\documentclass[seceq,supplement]{ptptex}

\usepackage{graphicx}
\usepackage{wrapft}

\newcommand{\be}{\begin{equation}}
\newcommand{\ee}{\end{equation}\noindent}
\newcommand{\bea}{\begin{eqnarray}}
\newcommand{\eea}{\end{eqnarray}}

\newcommand{\lapprox}{\raisebox{-0.5ex}{$\ 
\stackrel{\textstyle<}{\textstyle\sim}\ $}}
\newcommand{\maprightb}[1]{\smash{\mathop{
\hbox to 1cm{\rightarrowfill}}\limits_{#1}}}
\title{
Mesons at finite baryon density in $(2+1)d$
}

\author{
Costas G. \textsc{Strouthos} 
}

\inst{
\it Department of Physics, Duke University,  Durham, NC 27708-0305, USA\\
}

\abst{
We discuss the critical properies of the three-dimensional Gross-Neveu model 
at nonzero temperature and nonzero chemical potential. We also present 
numerical and analytical results for the in-medium interaction due to scalar
meson exchange. Further, we discuss in-medium modifications of mesonic
dispersion relations and wavefunctions. 
}

\begin{document}

\maketitle

\section{Introduction}\label{sec-intro}
Chiral phase transitions and the spectrum of excitations in 
strongly interacting matter at finite baryon chemical potential 
remain interesting challenges. 
Strongly interacting systems are intrinsically non-perturbative
and therefore most of our knowledge about the relevant phenomena comes from 
lattice simulations. Unfortunately, the complex nature of the determinant 
of the QCD Dirac operator at finite chemical potential makes it impossible to use
standard simulation techniques to study Fermi surface phenomena in Euclidean simulations. 

In order to understand what ingredients might play a decisive role
in more complex systems such as gauge theories, we have studied the
simplest non-trivial model simulable with $\mu \neq 0$ using standard algorithms, 
namely the three-dimensional Gross-Neveu model (GNM$_3$). 
Its Lagrangian in Euclidean metric is written in terms of 
 $4N_f$-component spinors $\psi,
\bar\psi$ as 
\begin{equation}
{\cal L}=\bar\psi(\partial{\!\!\!/\,}+m)\psi-{g^2\over{2N_f}}(\bar\psi\psi)^2.
\end{equation}
In the chiral limit $m=0$ the model has a global Z$_2$ symmetry
$\psi\mapsto\gamma_5\psi$, $\bar\psi\mapsto-\bar\psi\gamma_5$.
At tree level, the field $\sigma$ has no dynamics; it is trully an
auxiliary field. However, it acquires dynamical content by dint of quantum 
effects arising from integrating out the fermions. The model is renormalizable in 
the $1/N_f$ expansion unlike in the loop expansion \cite{rosen91}. Apart from the obvious 
numerical advantages of working with a relatively simple model there are several 
other motivations for studying this model.
At $T=\mu=0$ for sufficiently strong coupling $g^2$, chiral symmetry is
spontaneously broken by a condensate $\langle\bar\psi\psi\rangle\not=0$ leading 
to a dynamically generated
fermion
mass gap given by $M_f=g^2\langle\bar\psi\psi\rangle\gg m$ in the large-$N_f$ approximation.
The spectrum of excitations contains both baryons and mesons, i.e. the elementary fermions
$f$ and the composite $f\bar{f}$ states. The critical coupling
$g_c^2$ at which the gap $M_f/\Lambda_{UV}\to0$,
defines an ultra-violet stable fixed point of
the renormalisation group at which an interacting continuum limit may be taken.
This picture has been verified at next-to-leading order in $1/N_f$
and by Monte
Carlo simulations with finite $N_f$ \cite{hands93}. 

In the next section we will discuss the issue of chiral symmetry restoration at nonzero 
fermion chemical potential. 
In section \ref{sec:section3} we will discuss Fermi surface phenomena, such as long range interparticle
interaction in medium. We will also present results for dispersion relations (section \ref{sec:section4} 
and wavefunctions (section \ref{sec:section5}) of
massless particle-hole excitations on or near the Fermi 
surface. A study of the dispersion relation of spin-$\frac{1}{2}$ excitations  around the 
Fermi surface has shown that there is no mass gap characteristic of a BCS instability \cite{hands03} 
and the values of the Fermi liquid parameters (Fermi momentum, Fermi velocity) are in good
agreement with the $1/N_f$ predictions. 
However, in a different study of a related model, the $(2+1)d$ NJL model, which has an
$SU(2)\otimes SU(2)$ chiral symmetry, it was shown that although the BCS instability is also absent 
the properies 
of the quasi-particle spectrum is determined by physics outside the scope of the 
$1/N_f$ expansion \cite{hands02}. 

\section{Chiral phase transitions}
The action of the NJL model remains real even after the introduction
of nonzero chemical potential $\mu$, which means we can study the
physics of the high density regime using standard Monte Carlo
techniques.
In the presence of a Fermi surface with Fermi momentum $p_F$
the creation of $f\bar{f}$ pairs
with zero net momentum is energetically suppressed, because one can only create
particles with $p>p_F$. So as the fermion number density $\eta(\mu)$ grows
the chiral symmetry breaking is suppressed.
The large-$N_f$ description of the $\mu \neq 0$ chiral phase transition
predicts a first order transition for $T=0$ and a continuous transition for
$T>0$ \cite{klimenko}. Furthermore, the critical value of the chemical potential $\mu_c$
is equal to the value of the fermion mass at $\mu=0$, which indicates that
materialization of the fermion itself drives the symmetry restoration
transition.
Interactions as expected decrease $\mu_c$ below the mean field result \cite{hands95}.
Work by Stephanov \cite{stephanov96} suggests that any non-zero density simulation which
incorporates a real path integral measure proportional to $\det(MM^{\dagger})$
is doomed to failure due to the formation of a light baryonic pion from a quark $q$
and a conjugate quark $q^c$.
The NJL model however, does not exhibit such a pathology, because the realization of the
Goldstone mechanism in this model is fundamentally different from that in QCD.
In the NJL model the Goldstone mechanism is realized by a pseudoscalar
channel pole formed from disconnected diagrams and the connected diagrams
yield a bound state of mass $\approx 2 m_f$ \cite{hands99}. This implies the absence
of a light $qq^c$ state.

As expected our simulations of GNM$_3$
with $N_f=4$ \cite{strouthos01} did not provide any evidence for the
existence of a nuclear liquid-gas transition at $\mu<\mu_c$.
It was shown in \cite{buballa} that in the standard four-fermi models there is no saturation density
for stable matter. In order to get the saturation features the authors of
\cite{buballa} introduced a chirally invariant scalar-vector interaction term which
cures the binding problem.
Furthermore, our results showed that the second order nature of the
$T \neq 0$, $\mu=0$ transition remains stable down to low $T$ and large
$\mu$.
\begin{figure}[hbt]
\begin{center}
\begin{minipage}{ 0.48\linewidth}
\includegraphics[width=1.0 \linewidth]{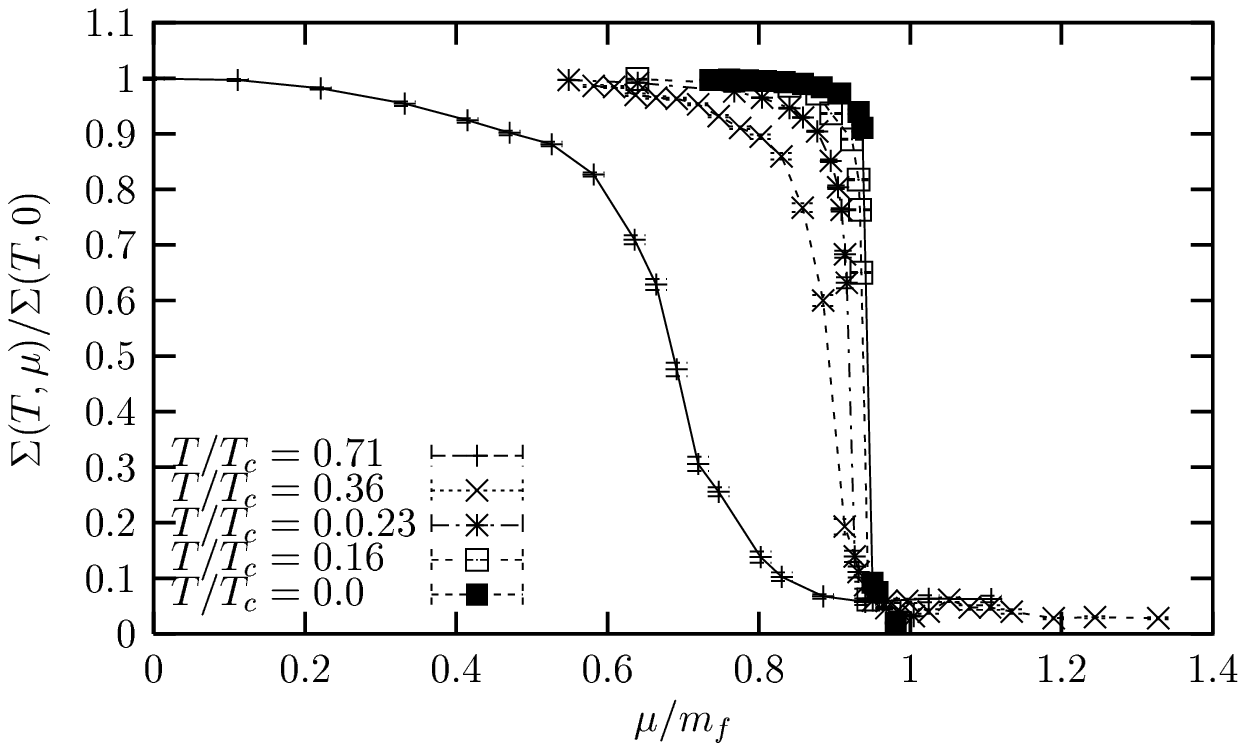}
\end{minipage}
\hspace{2mm}
\begin{minipage}{ 0.48\linewidth}
\includegraphics[width=1.0 \linewidth]{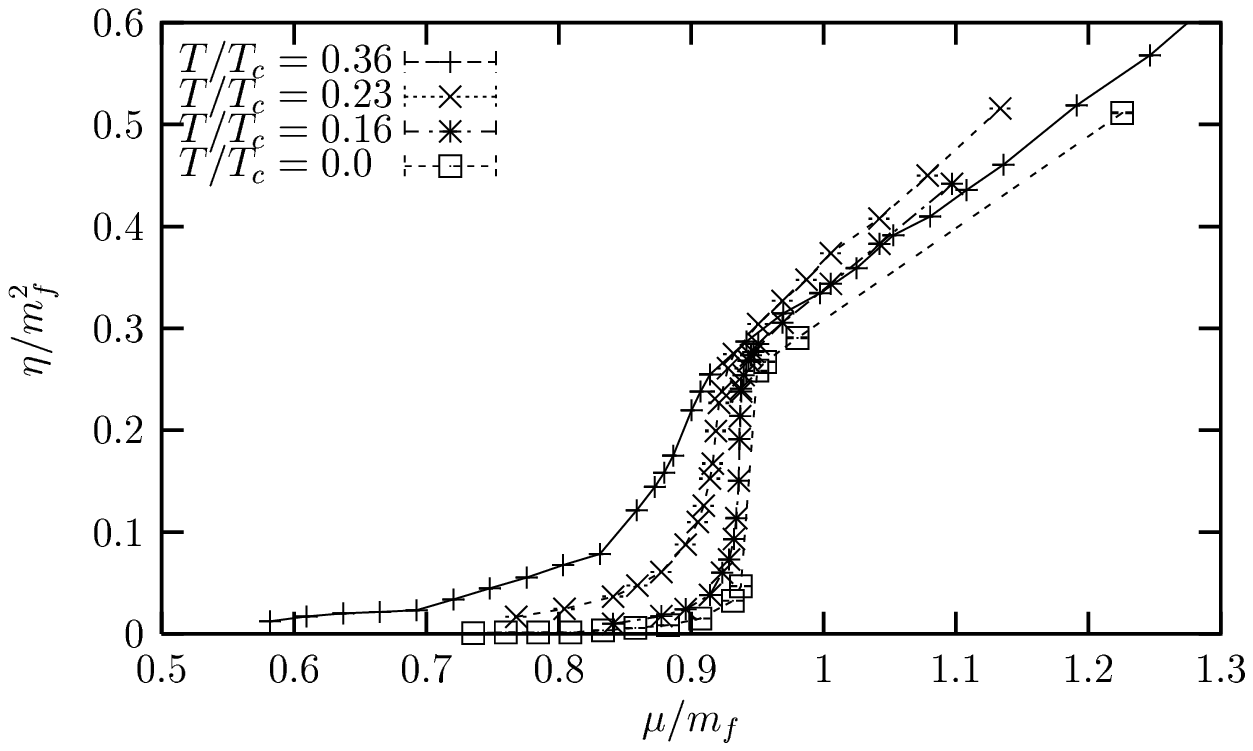}
\end{minipage}
\end{center}
\caption{
Chiral order parameter (left) and fermion number density (right) vs. chemical potential 
at various values of $T$.
}
\label{fig:1}
\end{figure}
In Fig.\ref{fig:1} (left) we plot the normalized order parameter $\Sigma(T,\mu)/\Sigma(T,0)$
($\Sigma\equiv\langle\sigma\rangle$)
as a function of $\mu/m_f$ at different values of $T$. It is clear from the shapes of these
curves that the transition becomes sharper as we decrease the temperature. In Fig.\ref{fig:1}
(right)
we plot the normalized fermion number density as a function of the chemical potential
at different values of $T$. In the limit $T \rightarrow 0$ we see that the fermion
density is strongly suppressed before the transition and then jumps discontinuously.
By performing detailed finite size scaling analysis
which allowed us to distinguish between second order and weak first order transitions
we showed that the tricritical point lies on the section
of the phase boundary defined by $T/T_c \leq0.23$, $\mu/\mu_c \geq0.97$ \cite{strouthos01}.
It was also shown some time ago \cite{kogut98} that the second order transitions 
belong to the $2d$ Ising universality class in accordance with the dimensional 
reduction scenario.

\section{The sigma propagator in medium}
\label{sec:section3}
In this section we examine the auxiliary scalar propagator $D_{\sigma}$ in the
presence of non-zero baryon density.
Although at tree level $\sigma$ is non-propagating, the leading order
$1/N_f$ expansion at $\mu=0$ predicts that it acquires dynamics through
quantum corrections due to virtual $q\bar{q}$ pairs, resulting in a propagator
of the form \cite{rosen91, hands93}
\begin{equation}
D_\sigma(k^2)={1\over
g^2}{{(4\pi)^{d\over2}}\over{2\Gamma(2-{\textstyle{d\over2}})}}
{M_f^{4-d}\over{(k^2+4M_f^2)F(1,2-{\textstyle{d\over2}};{3\over2};
-{k^2\over{4M_f^2}})}}.
\label{eq:Dsigma}
\end{equation}
Immediately we see the difference between this model
and QCD.
For $k^2\ll M_f^2$ the hypergeometric function $F\approx1$, implying that
to this order the $\sigma$ resembles a weakly-bound meson of mass
$M_\sigma=2M_f$; however,
the hypergeometric function
in the denominator indicates a strongly interacting $q\bar q$
continuum immediately above the threshold $2M_f$.
This implies that if truly bound, its binding energy is
$O(1/N_f)$ at best
(to our knowledge there have so far been no analytic calculations), implying
little if any separation between pole and threshold. A recent study using the maximum entropy 
method has shown evidence for a nonzero binding energy in the $\sigma$ channel for 
finite $N_f=4$ \cite{allton02}.
Since all residual interactions
are subleading in $1/N_f$, we surmise that all other mesons
are similarly weakly bound states of massive fermions, and hence effectively
described by a two-dimensional ``non-relativistic quark model''.

\begin{figure}[]
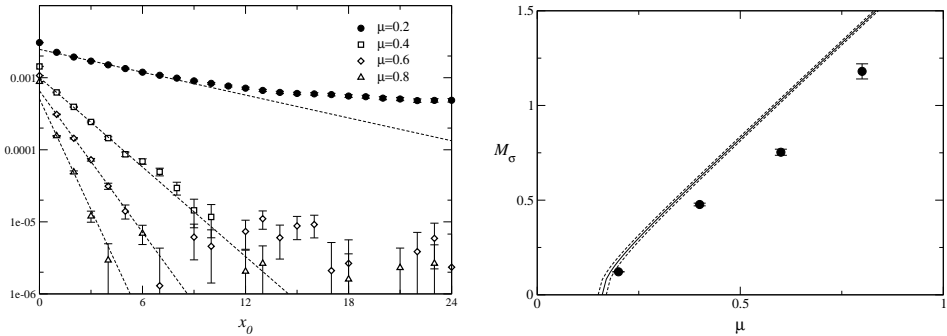

\begin{center}
\begin{minipage}{ 0.43\linewidth}
\includegraphics[width=1.0 \linewidth]{./sigma_corr_mon.eps}
\end{minipage}
\hspace{2mm}
\begin{minipage}{ 0.43\linewidth}
\includegraphics[width=1.0 \linewidth]{./msigma_mon.eps}
\end{minipage}
\end{center}
\caption{
The zero momentum $\sigma$ timeslice correlator for 4 values of $\mu$ (left)
and the $M_{\sigma}$ as a function of $\mu$ (right); the line is the leading order 
$1/N_f$ prediction.
}
\label{fig:2}
\end{figure}

Similarly, at nonzero chemical potential $\mu$ to leading order in the $1/N_f$
expansion, we have 
\begin{eqnarray}
D_\sigma^{-1}(k)&=&1-\Pi(k;\mu)\nonumber\\
&=&g^2\left[{1\over\Sigma_0}\int_{q,\mu=0}
\mbox{tr}{1\over{iq{\!\!\! /\,}+\Sigma_0}}
+\int_{q,\mu>\mu_c}\mbox{tr}{1\over{iq{\!\!\! /\,}+\mu\gamma_0}}
{1\over{i(q{\!\!\!/\,}-k{\!\!\!/\,})+\mu\gamma_0}}
\right],
\end{eqnarray}
where $\Pi$ is the virtual fermion-antifermion vacuum polarization bubble.
We have used the gap equation at $\mu=0$ to express $1/g^2$ in terms of the zero
density gap $\Sigma_0$ and assumed that the gap vanished in the
integral defining $\Pi$ for $\mu>\mu_c$.
The functional form of the propagator in the Hard Dense Loop (HDL) approximation
($k_0, |\vec{k}| \ll \mu$) in $d+1$ dimensions is given by \cite{hands03} 
\begin{eqnarray}
D_\sigma^{-1}(k_0,\vec k)=
{{g^2\mu^{d-3}}\over{(4\pi)^{d\over2}\Gamma(\textstyle{d\over2})(3-d)}}
\Biggl[4{{(3-d)}\over{(d-1)}}\mu^{3-d}(\mu^{d-1}-\mu_c^{d-1})+\;\;\;\;\;\;\;
\;\;\;\;\;\;\;\;\;\;\;\;\;\;\;\;\;\;\;\;\;\;\label{eq_Dsigma}\\
k_0^2+\vert\vec k\vert^2
\biggl(1-{{(3-d)(k_0^2+\vert\vec k\vert^2)^{1\over2}}\over
{k_0+(k_0^2+\vert\vec
k\vert^2)^{1\over2}}}\biggr)
+i{{k_0\vert\vec k\vert^2}\over{4\mu}}
\biggl(1+{{\vert\vec k\vert^2}\over
{[k_0+(k_0^2+\vert\vec
k\vert^2)^{1\over2}]^2}}\biggr)
+O\Bigl({k^4\over\mu^2}\Bigr)\Biggr].\nonumber
\label{htl}
\end{eqnarray}
An interesting observation is that in the static limit $k_0=0$
the momentum dependence of eq.~(\ref{htl}) vanished for the value $d=2$
implying complete screening for the static potential due to $\sigma$
exchange. Furthermore, for $\vec{k}=\vec{0}$ $D_{\sigma}$ is proportional
to a conventional zero-momentum boson propagator with mass
\begin{equation}
M_\sigma=2\sqrt{\mu(\mu-\mu_c)},
\label{eq:msigma}
\end{equation}
implying that just above the transition there is a tightly bound state.
In Fig.~\ref{fig:2} we show the results for the sigma correlator obtained 
from numerical simulations with $N_f=4$ and $1/g^2=0.75$ on 
$48 \times 32^2$ lattices.  The value of the critical chemical potential $\mu_c=0.17(1)$. 
\smallskip
\bigskip
The numerical results for $M_{\sigma}$ fall $20-30\%$ lower than the leading order 
prediction, which is consistent with a correction of $O(N_f^{-1})$.

\section{Mesonic dispersion relations}
\label{sec:section4}
In this section we investigate mesonic states by measuring the connected contribution 
to the correlation functions 
${\cal C}_\Gamma(\vec k,x_0)=\sum_{\vec x}\langle j_\Gamma(\vec 0,0)
j_\Gamma^\dagger(\vec
x,x_0)\rangle e^{-i\vec k.\vec x}$ where the bilinears $j_\Gamma(x)$
are defined with
scalar, pseudoscalar or vector quantum numbers.
In terms of staggered fermion fields $\chi,\bar\chi$ the operators are
\begin{equation}
j_{\bf 1}(x)=\bar\chi_x\chi_x;\;\; 
j_{\gamma_5}(x)=\varepsilon_x\bar\chi_x\chi_x;\;\;
j_{\gamma_i}(x)={\eta_{ix}\over2}[\bar\chi_x\chi_{x+\hat\imath}+
                          \bar\chi_{x+\hat\imath}\chi_x],
\end{equation}
where $\eta_{1x}=(-1)^{x_0}$, $\eta_{2x}=(-1)^{x_0+x_1}$ and
$\varepsilon_x=(-1)^{x_0+x_1+x_2}$. As before we study the timeslice
correlators in each channel as a function of spatial momentum $\vec
k\parallel\hat x$.
\bigskip
\begin{figure}[hbt]
\begin{center}
\begin{minipage}{ 0.53\linewidth}
\includegraphics[width=1.2 \linewidth]{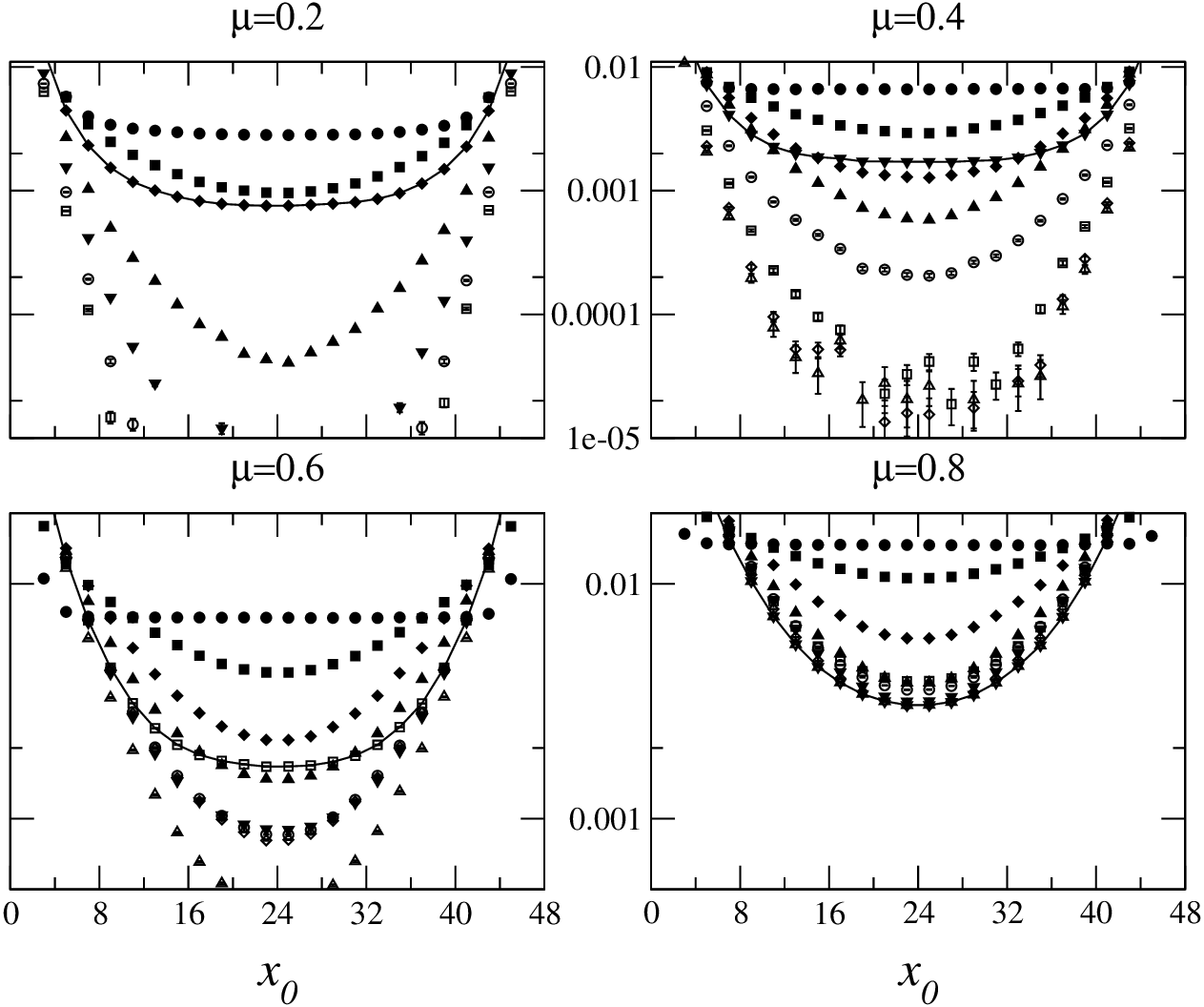}
\end{minipage}
\end{center}
\caption{
Pseudoscalar correlator ${\cal C}_{\gamma_5}(\vert\vec k\vert,x_0)$
at 4 different $\mu$
for momenta $\vert\vec k\vert=0$ (filled circles),
${\pi\over16}$ (filled squares), ${\pi\over8}$ (filled diamonds),
${3\pi\over16}$ (filled up triangles), ${\pi\over4}$ (filled down triangles),
${5\pi\over16}$ (open circles), ${3\pi\over8}$ (open squares), ${7\pi\over16}$
(open diamonds) and  ${\pi\over2}$ (open up triangles).
}
\label{fig:psdisperse}
\end{figure}
It was shown \cite{hands03} that in free field theory for $\mu>\mu_c$ the the $x_0 \rightarrow 0$ 
limit of ${\cal C}_\Gamma(\vec k,x_0)$
is dominated by a continuum of particle-hole pairs at or near the Fermi surface, which effectively 
cost zero energy to excite. 
The generic result is that for
$\vert\vec k\vert\leq2\mu$ the decay is algebraic, with
\begin{equation}
{\cal C}_\Gamma(\vec k,x_0)\propto \left\{
\begin{array}{ll}
x_0^{-2}         & \mbox{$\vert\vec k\vert\ll\mu$} \\
x_0^{-{3\over2}} & \mbox{$\vert\vec k\vert\simeq2\mu$}.
\end{array}
\right.  
\end{equation}
Only once $\vert\vec k\vert>2\mu$ does it become kinematically impossible to
excite a pair with zero energy, resulting in exponential decay:
\begin{equation}
{\cal C}_\Gamma(\vec k,x_0)\propto x_0^{-{3\over2}}\exp\Bigl(-(\vert\vec
k\vert-2\mu)x_0\Bigr).
\end{equation}
The sequence of plots in Fig.~\ref{fig:psdisperse} is in qualitative agreement
with these findings. For each $\mu$ there is a particular value of
$\vert\vec k\vert$, highlighted with a solid line in the plots, for which the
temporal falloff is particularly slow, corresponding to
$\vert\vec k\vert\approx2\mu$ (e.g. for $\mu=0.4$ the slow falloff
occurs for $\vert\vec k\vert={\pi\over4}\simeq0.785$).
For $\vert\vec k\vert$
larger than this value the decay
is much steeper, although only for $\mu=0.2$ does it resemble an
exponential form.
Because of the technical difficulties in treating correlators with power-law
decays in a finite volume we have made no attempt to
fit the numerical data for ${\cal C}_{\gamma_5}(\vert\vec k\vert,x_0)$
to a functional form.

\begin{figure}[]
\begin{center}
\begin{minipage}{ 0.69\linewidth}
\includegraphics[width=1.0 \linewidth]{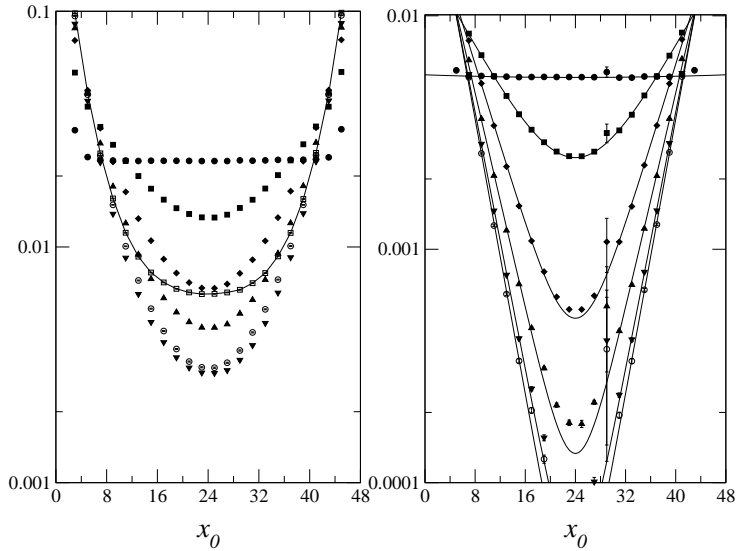}
\end{minipage}
\end{center} 
\caption{Vector correlators ${\cal C}_{\gamma_\perp}(\vert\vec k\vert,x_0)$
(left) and
${\cal C}_{\gamma_\parallel}(\pi-\vert\vec k\vert,x_0)$ (right) at $\mu=0.6$.
The symbols have the same
meaning as in Fig.~\ref{fig:psdisperse}.}
\label{fig:rhodisp}
\end{figure}

Light states in the $\rho$ channel have long been of interest.
Here we discuss tentative results in the vector channel, corresponding 
to the quantum numbers of the $\rho$ meson. 
In this case for $\vert\vec k\vert>0$
we can distinguish
between ${\cal C}_{\gamma_\parallel}(\vec k)$,
in which the component of the vector is parallel
to $\vec k$, and ${\cal C}_{\gamma_\perp}(\vec k)$.
In Fig.~\ref{fig:rhodisp} we plot
the correlator for several $\vert\vec k\vert$ values at $\mu=0.6$ in each case.
For ${\cal C}_{\gamma_\perp}$
the curves are qualitatively very similar to those of
Fig.~\ref{fig:psdisperse} at $\mu=0.6$, with a distinguished momentum
$\vert\vec k\vert={3\pi\over8}$.
The correlator ${\cal C}_{\gamma_\parallel}$ is much smaller in
magnitude, and is consistent with exponential rather than algebraic decay. The
lines are fits of the form ${\cal C}_{\gamma_\parallel}(\vert\vec k\vert,x_0)=
A(e^{-Ex_0}+e^{-E(L_t-x_0)})$.
For small $\vert\vec k\vert$ the resulting $E(\vert\vec k\vert)$ 
resembles that of a massless
pole, ie. $E=\beta_0\vert\vec k\vert$, with velocity $\beta_0\approx0.5$.
It should be mentioned, however, that although ${\cal C}_{\gamma_\parallel}$ and
${\cal C}_{\gamma_\perp}$ still differed, no evidence for a massless pole
was seen in the data at $\mu=0.8$.
A potential problem for this picture is that the extracted velocity
$\beta_0\approx0.5$ is less than
$\beta_F\approx1$, implying that quasiparticles would experience damping via
\v Cerenkov radiation of zero sound. Clearly further work exploring the
systematic
effects of varying $\mu$, $g^2$, $\vec k$ and volume
will be needed for a more complete understanding to emerge.
\smallskip
\bigskip

\section{Mesonic wavefunctions and Friedel oscillations}
\label{sec:section5}
In this section we expose another characteristic of the Fermi surface 
by studying the spatial correlations between 
fermions and antifermions in various mesonic channels, probed via the
wavefunction $\Psi(\vec x)$ defined by 
\begin{equation} 
\Psi_\Gamma(\vec x)=\lim_{x_0\to\infty}\Psi_\Gamma^{-1}(\vec x=\vec0) 
\sum_{\vec y}\langle{\cal G}_q(\vec 0,0;\vec
y,x_0)\Gamma{\cal G}_{\bar q}(\vec 0,0;\vec y+\vec x,x_0)\Gamma\rangle,
\end{equation}
where as usual 
$\Gamma$ projects out the quantum numbers of the channel of interest.
In the GNM$_3$ and related models $\Psi(\vec x)$ is technically 
much easier to define and measure than in QCD-like theories where they are 
gauge-dependent.
Meson wavefunctions in the GNM$_3$ model have been studied at $T,\mu=0$ in
\cite{strouthos02}, where further technical details are given.

It is easy to show \cite{hands03} that in the large-$N_f$ approximation the PS wavefunction in 
the chirally broken phase is given by 
\begin{equation}
\lim_{x_0\to\infty}C(\vec y;x_0)\sim {M\over x_0}e^{-2Mx_0}\exp\left(-
{{\vert\vec y\vert^2M}\over4x_0}\right).
\end{equation}
The general profile of the wavefunction is a gaussian with width increasing as
$\surd x_0$.
In the chirally symmetric phase we get, 
\begin{equation}
\lim_{x_0\to0}C(\vec y;x_0)\sim{\mu\over x_0}e^{-2\mu x_0}J_0(\mu\vert\vec
y\vert).
\label{eq:friedel}
\end{equation}
The wavefunction profile no longer changes with  $x_0$, but instead oscillates
with a spatial frequency determined by $\mu$, which to this order may be
identified with the Fermi momentum $k_F$. The oscillations observed in $C(\vec y;x_0)$ 
are characteristic
of a sharp Fermi surface and are reminiscent of oscillations in
either the density-density correlation function, or the screened inter-particle
potential, in degenerate systems generically known as {\em Friedel
Oscillations\/} \cite{fri}.
\bigskip
\smallskip
\begin{figure}[hbt]
\begin{center}
\begin{minipage}{ 0.46\linewidth}
\includegraphics[width=1.0 \linewidth]{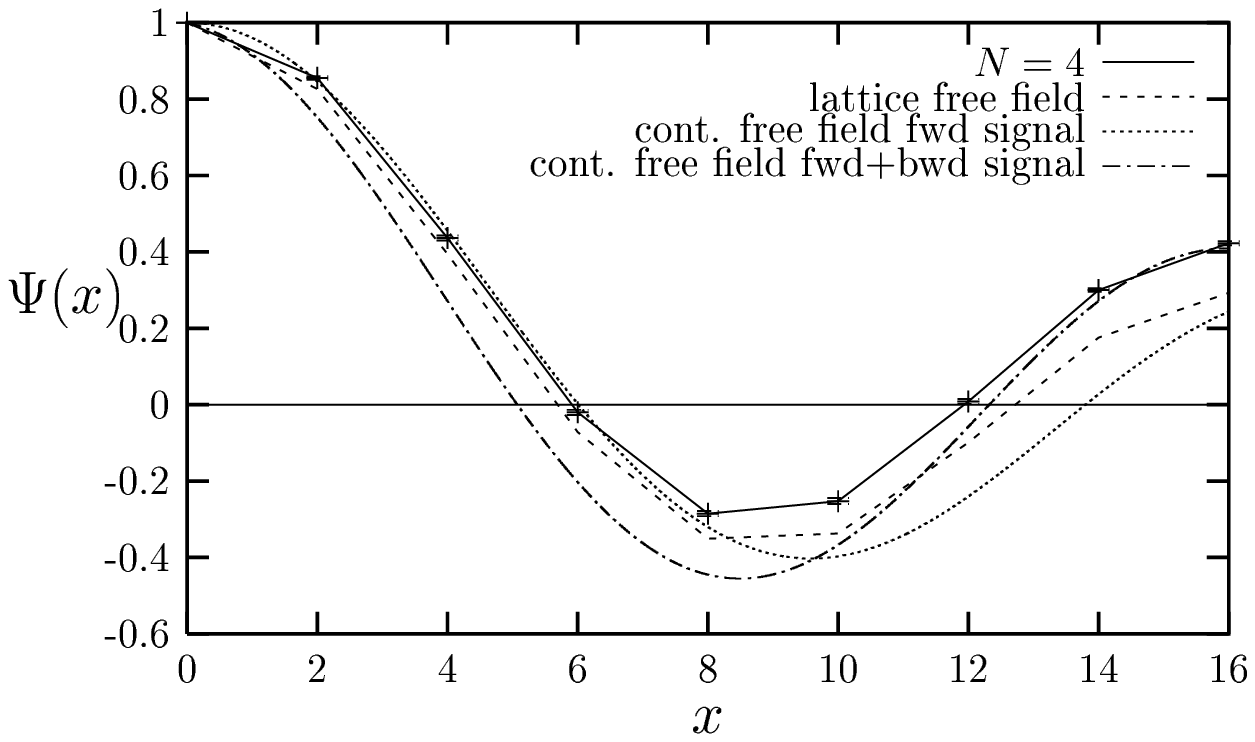}
\end{minipage}
\hspace{2mm}
\begin{minipage}{ 0.46\linewidth}
\includegraphics[width=1.0 \linewidth]{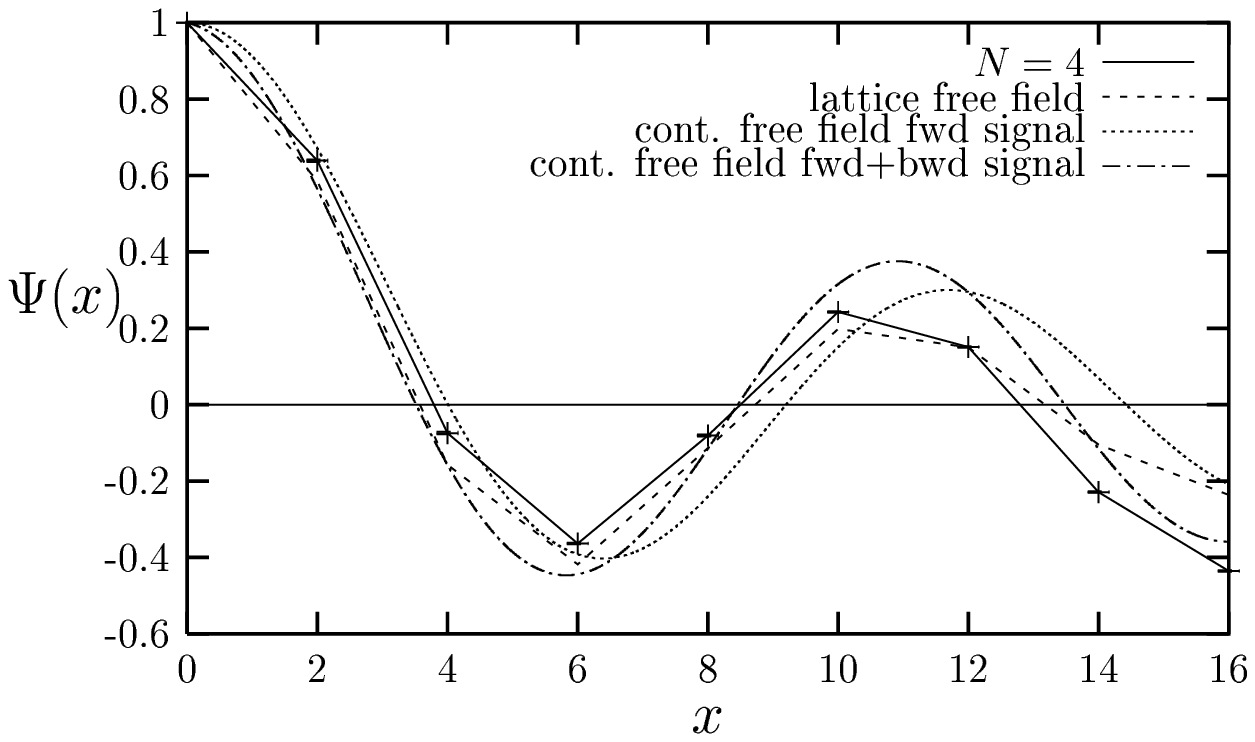}
\end{minipage}
\end{center}
\caption{
Scalar wavefunction at $\mu=0.4$ (left) and $\mu=0.6$ (right).
}
\label{wvfn}
\end{figure}
In Fig.~\ref{wvfn} we plot the PS wavefunctions at $1/g^2=0.75$ for 
$\mu=0.4, 0.6$. The simulation data are connected by solid lines. As $\mu$, 
and hence $k_F$ increases the oscillations decrease in wavelength in accordance 
with the theoretical expectation and provide a graphic illustration of the
presence of a sharp Fermi surface. The dashed lines show measurements taken with the same
lattice parameters but with the
interaction switched off. The disparity with the interacting theory is
small, though increasing with $\vert\vec x\vert$,
showing that the free field description of
the oscillations is qualitatively correct. A possible explanation
is the infinite value predicted for the
Debye mass in section \ref{sec:section3}, implying that interactions between static
quarks at non-zero $\vert\vec x\vert$ are completely screened.
Also shown is the
theoretical form (eq.~\ref{eq:friedel})
$\Psi(x)=J_0(k_F x)$ with $k_F\equiv\mu$, showing good agreement with the data
for small $x\lapprox2k_F^{-1}$. Unfortunately it appears hard to
obtain more quantitative information, such as an independent fit for
the Fermi momentum $k_F$, because $\vert J_0(kx)\vert$
decays only as $x^{-{1\over2}}$. This means that fits should
not only include the backwards-propagating signal $J_0(k_F(L_s-x))$ but also
image contributions $J_0(k_F(nL_s-x))$ -- our attempts to find a
satisfactory fit were unsuccessful. The figures therefore
simply show both the ``forwards'' $J_0(\mu x)$
and ``forwards-and-backwards'' $J_0(\mu x) + J_0(\mu(L_s-x))$ forms, showing
that neither gives a satisfactory description of the data over the full range
of $x$ and $\mu$.
We also checked that there are no significant differences among
the various
mesonic channels, implying that in contrast to the situation at $\mu=0$
\cite{strouthos02}, effects due to eg. $\sigma$ exchange are very hard to detect.

\section{Summary}
The four-fermion models remain the only interacting fermionic field 
theories both simulable by standard lattice methods at $\mu \neq 0$ 
and displaying a Fermi surface, thought to be an essential figure
of dense quark matter. We have shown that the $\mu \neq 0$, $T=0$ 
transition is strongly first order and a tricritical point exists 
in the $(\mu,T)$ plane at very low $T$. 

We also discussed in-medium effects on the character of mesons. 
The large-$N_f$ auxiliary boson propagator has a branch-cut in the 
complex-$k$ plane which is modified to an isolated pole in the high 
density phase. Lattice simulations with $N_f=4$ have verified this 
with reasonable precision. We have also shown that the connected 
(flavor non-singlet) diagrams instead of showing exponential falloff
with Euclidean time, they generically decay algebraically, signalling 
the presence of massless particle-hole excitations. 
We also presented tentative evidence for a massless pole in the 
vector channel, which is possibly a manifestation of zero sound.
Furthermore, 
a graphic confirmation of the presence of a sharp Fermi surface
was shown by the oscillatory behavior in mesonic wavefunctions, 
which resemble the Friedel oscillations in many-body physics. 

\section*{Acknowledgements}
The work presented here was done in collaboration with Simon Hands \cite{hands03}, 
John Kogut \cite{strouthos01, hands03}
and Thao Tran \cite{hands03}.

\end{document}